# Modelling information manufacturing systems

## Thanh Thoa Pham Thi* and Markus Helfert

School of computing, Dublin City University,
Glasnevin, Dublin 9, Ireland
Fax: +353 1 700 5442
E-mail: Thoa.Pham@computing.dcu.ie
E-mail: Markus.Helfert@computing.dcu.ie
*Corresponding author

Abstract: The manufacture of an information product (IP) is akin to the manufacture of a physical product. Current approaches to model such information manufacturing systems (IMS), lack the ability to systematically represent the dynamic changes involved in manufacturing (or creating) an IP. They also have limitations to consistently include aspects of process and information management at an organisational level. This paper aims to address these limitations and presents a modelling approach, the IASDO model. Our work also represents a framework to evaluate the quality of the meta-models for IMS modelling which enable us to compare the IASDO model with current approaches.

Keywords: information quality; meta-model; information system architecture; dynamic specialisation.

Biographical notes: Thanh Thoa Pham Thi is currently a Postdoctoral Researcher at Dublin City University. She received her PhD Degree in Information Systems from the University of Geneva, Switzerland in 2005. Her research interests include information systems engineering and information quality.

Markus Helfert is a Lecturer in Information Systems at Dublin City University. He received his PhD Degree in Information Management from the University of St. Gallen, Switzerland and a Master's Degree in Business Informatics from the University Mannheim, Germany. His research is centred on information quality management and includes research areas such as data warehousing, business process management, healthcare information systems and supply chain management.

## 1 Introduction

Information quality impacts on business and decision-making, especially information quality in computer-based systems, which are becoming critical to many organisations. Over the last decade, academics addressed challenges in information quality and have proposed several methods to measure and improve information quality in such systems

.

(Eppler et al., 2004). One typical method is to manage information as an IP and to view the manufacture of an IP as a sequence of processes that must be represented accurately (Ballou et al., 1998; Shankaranarayanan et al., 2000). In this observation, manufacturing an IP is akin to the manufacture of a physical product in a supply chain network. Raw materials, storage, assembly, processing, inspection, rework and packaging (formatting) are all applicable to an IP.

An Information Manufacturing System (IMS) (Ballou et al., 1998) is an information system that produces predefined IPs. In order to describe such systems, it is common to describe the required system elements, their structure and the process dynamics through various models. These are commonly referred to as enterprise models or information system architectures (Bernus et al., 1998). An essential element in manufacturing an IP is the underlying processes. In addition, information and organisational aspects should not be neglected, as processes in information manufacturing may need organisational responsibilities for execution. However, information-manufacturing processes are often seen as far more complex to manage than industrial manufacturing processes.

In order to develop and maintain a consistent and adequate (enterprise) model, user requirements and dynamic changes in the enterprise environment have to be captured. Similar to the requirements of information system architectures, models of information manufacturing systems should include at least three aspects:

- static aspects that concerns information and information structure in form of raw data, component data or IP

- dynamic aspects that concerns information processing

- organisational aspects that concern the management of the information system at an organisational level.

Although attempts have been made to provide models for information manufacturing systems, current approaches have their drawbacks. Indeed, certain critical aspects of the stages that an IP goes through within the manufacturing system have not been addressed. Most current approaches lack the ability to systematically represent the dynamic changes involved in manufacturing an IP and to represent process and information management at an organisational level. Besides, they often do not take into account the interrelation between these aspects to ensure a consistent specification.

In order to address some limitations of the current approaches, in this paper we propose an object-oriented model for IMS modelling; the Integrated Aspects of Static, Dynamic and Organisation model (IASDO). The IASDO model allows us to describe the manufacture of IPs by modelling the structure of IPs, the information processing and organisational roles for information management. In addition, our approach allows us to consider the interrelation between these aspects. By integrating these aspects, the IASDO model ensures an accurate and consistent specification.

The reminding sections of this paper are organised as follows: Section 2 outlines some related work on model quality and presents a framework for evaluating the quality of meta-models. Section 3 provides an overview of some selected methods and models for IMS modelling and an illustration of these in a library management scenario. Section 4 aims to describe the IASDO model, including its concepts and rules. Section 5 applies the IASDO model to the library management scenario. Section 6 aims to evaluate the IASDO model and the models mentioned in Section 2, followed by a critical review of the IASDO model. Section 7 concludes our work and gives some indication for further research.

## 2 Model quality and meta-model quality

Information manufacturing systems are defined as information systems that produce pre-defined IPs (Ballou et al., 1998). Normally, a system is described by following some form of a System Life Cycle (SLC) that includes various analysis and design stages. The result of this modelling process is a conceptual model, which can be described by an explicit or implicit meta-model. In addition, the manner in which the activities are conducted at the various stages of the SLC influence the quality of the deliverables at each stage. Consequently these activities influence the quality of the eventual system (Duggan and Reichgelt, 2006). As a result, the modelling process and the conceptual model impact on the overall system quality.

Several researchers have addressed the quality evaluation of conceptual models. One of the common definitions for conceptual model quality is

> "the total quality of features and characteristics of a conceptual model that bear on its ability to satisfy stated or implied needs." (Moody, 2005)

Based on this definition, and in order to develop a framework for conceptual model quality, Moody (2005) synthesised eight different approaches (deductive, codification, inductive, social, analytical, reverse inference, Goal-Question-Metric model and Dromey's methodology). In conclusion, they stated that there is no common standard for conceptual model quality. Also Schuette and Rotthowe (1998) has proposed guidelines for modelling, which include principles that can be used to evaluate the quality of models. In order to reduce the subjectivism of designers, they have proposed a standard of information modelling. For instance, concerning the principle of language adequacy, they stated the following: a model is complete according to the meta-model, if the relationships between the information objects described in the meta-model are applied in the model itself in the same way. Moody and Shanks (1994), have proposed a quality framework for quality evaluation of data models. This framework is composed of six quality factors: completeness, integrity, understand-ability, simplicity, integration and implement ability. Each factor is associated with a weight, which indicates the importance of the factor. Furthermore, Lindland et al. (1994) and Krogstie et al. (1995) proposed a quality framework that focuses on three main criteria: syntax, semantic and pragmatics. Syntactical quality of a model relates to the model-completeness compared to the meta-model, as well as the consistency of the model compared to the meta-model. Semantic quality depends on the relevance of models for the modelling domain (modelling subject). The pragmatic quality of a model depends on the easy comprehension of this model and the feasibility concepts.

In the information quality research area, the critical objective is to improve the quality of data and information. Data quality and information quality (both terms are interchangeable in this paper) have been defined by widely accepted dimensions such as completeness, correctness, consistency and timeliness (Pipino et al., 2002). Following this approach, data are described by a conceptual model. Completeness and consistency of data may be dependent on integrity rules defined on the conceptual model. Consequently, the conceptual model may impact on the quality of data.

8      T.T. Pham Thi and M. Helfert

At a different level, the researches mentioned above have presented relationships between the meta-model and model quality. However, they have not described how the meta-model impacts on the model quality for example, the expressive power of the meta-model impacts on the expressive power of the conceptual model; whether the meta-model includes explicit rules for validating the conceptual model or validate the model implicitly by the experience of its designers. Therefore a meta-model of high quality may help to improve or ensure the quality of the conceptual model, therefore improving data quality (Figure 1).

Figure 1    Relations and influences on information quality between different model levels

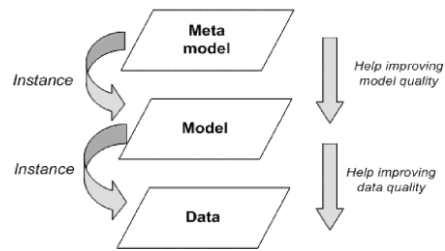

Certainly, the quality of a meta-model depends on how it supports designers in the modelling process. Generally, following quality criteria for meta-models is important. Criteria concerning subjective evaluation or social method evaluation (i.e., interview) such as simplicity and understand-ability are beyond the scope of this paper.

- Completeness: the meta-model allows us to describe different information that can not be described (or limited) by other meta-models

- Consistency: the meta-model is consistent itself and helps designers to obtain a consistent model (for instance by proposing validation rules)

- Accuracy: the meta-model is accurate in itself and helps designers to obtain an accurate model (for instance by proposing validation rules).

These criteria are used to evaluate the quality of meta-models in the following sections.

## 3 Related works

This section aims to present an overview of current methods and models used for IMS modelling. In fact, these are meta-models, however in the remainder of this paper, for simplicity, we also use the term 'model'.

Modelling of Information Manufacturing Systems has attracted several researchers over the last decade. Different methodologies and models have been proposed to model the process by which the IP is manufactured such as IP-UML (Scannapieco et al., 2005) or IP-Map (Ballou et al., 1998; Shankaranarayanan et al., 2000). In addition, some other approaches related to IMS modelling have being proposed in a broader sense. In this paper, we selected two such broader approaches, Data Flow Diagrams (DFD) (Demarco, 1978) and Event Driven Process Chains (EPC) (Keller et al., 1992).

IP-Map focuses on representing details associated with the manufacture of an IP. The model uses graphical constructs to describe processes, input and output information of processes, customer receiving output information, decisions, IS boundary, organisation boundary and department/role concept associated with each construct representing organisational responsibilities. IP-Maps are typically used to describe the composition of one IP and thus, the relations between several IPs are often not considered. Besides, they do not describe dynamic changes of information.

An adaptation of the IP-Map concept is the IP-UML approach, which combines IP-Map and UML. Scannapieco et al. (2005) propose a data analysis model, a quality analysis model and a quality design model. The data analysis model aims to identify interesting data and their composition by proposing stereotypical classes such as raw data, component data and IP. An IP is composed of raw data or component data or both. The quality model identifies quality requirements on each data item specified in the data analysis model. The quality design model combines the IP-Map and activity diagram of UML. It also describes data and processes in order to verify the satisfaction of quality requirement. The model typically does not represent relations between different information manufacturing processes. As such, it only represents a local description of the manufacture of an IP. Furthermore, neither dynamic changes of information nor the privileges of organisational units for information access have been described.

DFD, have been used for many years. They focus on describing processes as well as their inputs and outputs. External entities and data stores are also included in the model. External entities may send/receive data to processes. Due to the focus on processes, relations between data are not described and data changes are not captured. The model does not take into account organisational structures, so neither responsibility for processes nor control of data access is mentioned. However, the model includes a set of rules to evaluate the correctness of a DFD (Celko, 1987).

Event Driven Process Chains are typically used to describe business processes. They describe information exchanged and functions of the organisation. The model includes graphical constructs to model functions, events, input data, output data, organisational units and relations between these aspects. The model does not systematically take into consideration the relation between the aspects and does not ensure a consistent specification. Furthermore, the model does not mention control of data access.

In the following, we illustrate the IP-Map and EPC model as examples for two modelling approaches: IP oriented and business process oriented.

We illustrate the concept by modelling a library management application using the following simplified scenario. A library holds copies of books that may be borrowed by readers. Copies are units representing physical books that belong to one logical document. A logical document has one or many correspondent copies. Readers request a loan by indicating that they would require this document. The librarian can make a loan or a reservation for the reader depending on the status of a corresponding copy of the document. In the case of a reservation the librarian will notify the reader and record that this particular copy is 'blocked'. The reader can then borrow the blocked copy, which becomes a borrowed copy, or alternatively the reader can cancel the reservation and the copy is unblocked. The workflow is finished when the reader returns the copy or cancels the reservation.

Figure 2 illustrates a specification of the process creating a Loan using IP-Maps. Select document, Request a Loan, Create a loan, Create a reservation are processes. Loan and Reservation are IPs. Reader is a source of data as well as a consumer because he or she provides and uses information for the creation of IPs Loan and Reservation.

Figure 2   Process Create a loan for creating loan described with IP-Map

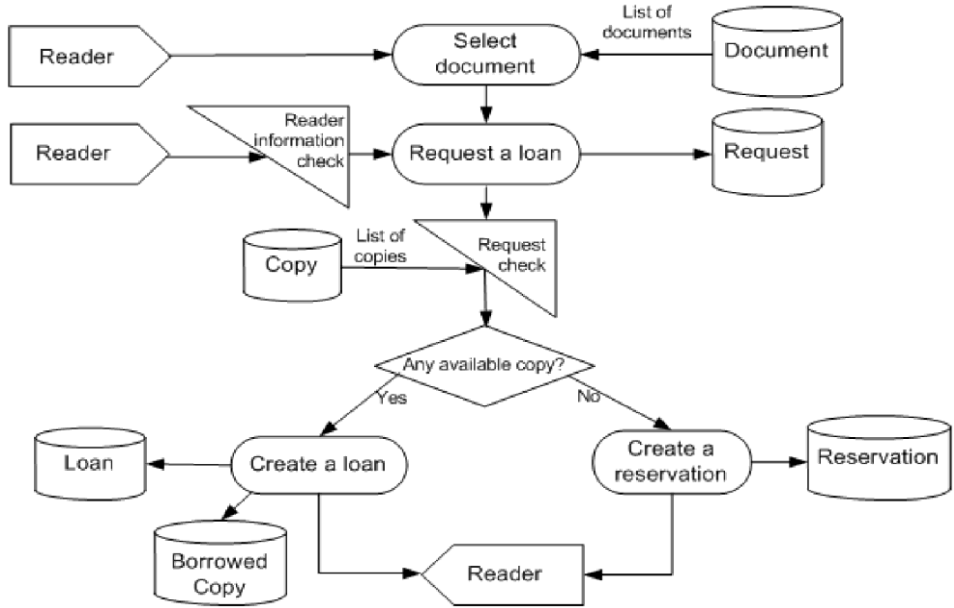

As a consequence of creating a loan, a borrowed copy is created. Using IP-Map, relations between Document and Copy; Copy and Borrowed Copy; Reservation, Loan and Request are not described. The designer may add integrity rules to ensure a consistent specification, for example that the borrowed copy of a loan belongs to the same document as the requested one. Because there is no relation between information (for example between Copy and Borrowed Copy), the specification does not describe how information changes, step by step through the manufacturing process.

Figure 3 presents a description of the business process of making a loan using the EPC modelling approach. Reader requests a loan, Copy available, No copy available are events. Librarian is an organisational unit who is responsible for the execution of processes: Make a request, Make a loan and Make a reservation. After the execution of processes Make a loan and Make a reservation, some data are modified. These include the change of state of request and copy or the creation of new data on request inventory and copy inventory. As illustrated in Figure 3, the specification lacks relations between various IPs. However, this would be required to ensure a consistent specification with precise specification and rules. Organisational concept is introduced in the model; it describes responsibility for the process execution; however, the security on information accessibility of organisational units is not mentioned.

An overview of some remarks of various modelling approaches mentioned above is provided in Table 1.

Figure 3   Business process Make a loan described with EPC

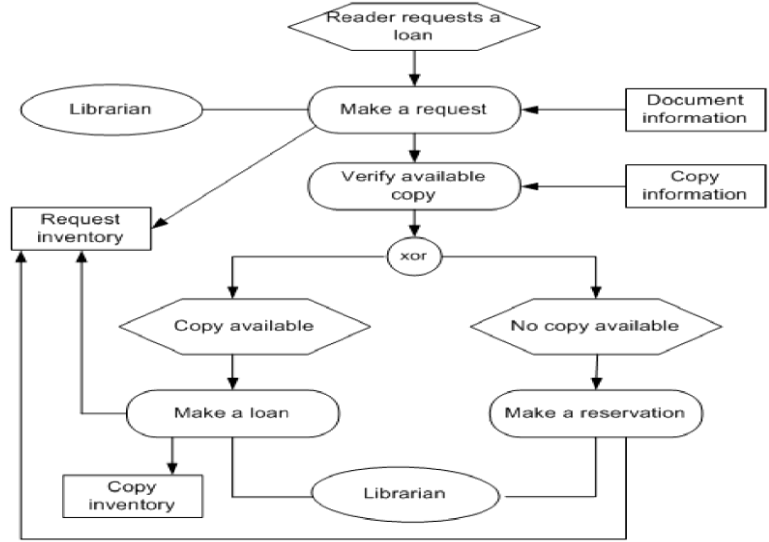

Table 1   Remarks on selected models/methods for IMS modelling

| Methodology/model | Main characteristic | Remarks |
| --- | --- | --- |
| IP-Map | Focus on manufacturing of one IP product | Do not describe dynamic changes of information |
|  |  | No relation between IPs |
|  |  | Do not take into account interrelations between three aspects |
| IP-UML | Focus on manufacturing one IP product | No relation between IPs |
|  |  | Do not describe dynamic changes of information |
|  |  | Privilege on data access is not mentioned |
|  |  | Do not take into account interrelations between three aspects |
| EPC | Focus on business process | No relation between data |
|  |  | Do not describe dynamic changes of data |
|  |  | Privilege on data access is not mentioned |
|  |  | Do not take into account interrelations between three aspects |
| Data Flow Diagram | Focus on information processing | No relation between data |
|  |  | Responsibility for process execution is not mentioned |
|  |  | Privilege on data access is not mentioned |
|  |  | Do not take into account interrelations between three aspects |

## 4   The IASDO model

In the following sections, we describe the basic concepts of the IASDO model (Pham Thi, 2005) that includes concepts of static, dynamic and organisational aspects of information manufacturing systems.

### 4.1   Static concepts

Static concepts present a static view of the system and include the structure of information and relations between information entities. Information may be raw data (material data), component data or IPs. The information elements are modelled using the class concept. The only type of relation between classes is Existential Dependency (ED), by which we are able to include more semantic information compared to similar concepts (e.g., Snoeck and Dedene, 1998). Our concept allows us to automatically validate rules concerning cycles in the schema (Pham Thi and Léonard, 2006). It also supports model evolution (Léonard, 1999).

One form of ED relation is the imperative ED relation. An imperative ED relation between a class C1 and a class C2 means that the existence of any object o1 of the class C1 depends on one and only one object o2 of the class C2. This dependency is permanent through the life of object o1. For example, there is an ED relation between a class called Copy and a class called Document. A copy can not exist if there is no linkage to a corresponding Document. The ED relation is transitive. In contrast, a non-imperative ED is an optional ED, i.e., an object o1 in class C1 depends on zero or one objects o2 in class C2.

Dynamic Specialisation (DS) is a particular case of an ED relation (Al-Jadir and Léonard, 1999; Pham Thi, 2005), which can be described by the following features: DS graph, active/inactive status, access-view and loops.

- DS graph. A DS graph is defined by a set of classes (nodes) linked by DS links
  (edges). If a class C1 is linked to a class C2 by a DS link, then C1 is a direct sub-class of C2 and C2 is a direct super-class of C1. For any object o1 (sub-object) of class C1, its existence depends on the existence of one and only one object o2 (super-object) of C2. A DS graph has one root class. An object can belong to several classes in the same DS graph. If an object belongs to a class C then it belongs to all super-classes of C (i.e., direct and indirect super-classes).

- Active/inactive status. An object that belongs to a class can be active or inactive in
  that class. Active is the default status. When an object changes from a class C2 to its sub-class C1, it may remain in C2 or leave C2. In the first case, the object remains active in the super-class C2, while in the second case the object becomes inactive in C2. This is indicated by the specialisation link. An inactive object cannot be updated, but can be consulted, i.e., one can get its attribute values.

- Access-view. An access-view of a class C consists of all the defined attributes and
  methods of class C, plus the selected properties of the super-classes of C. Thus the concept of access-view of C allows us to include or exclude properties of the super-classes of C. This differs from classical inheritance.

- Loop. An object can migrate from a class C to its sub-classes. In addition, there are
    some situations where an object may migrate from a sub-class C1 back to one of its
    super-classes. This presents an object migration loop, which requires keeping the
    trace of the object. Figure 4 illustrates such a situation; Copy is a super-class of
    AvailableCopy; AvailableCopy is a super-class of BorrowedCopy. This in turn is a
    super-class of Returned copy. By a loan, an AvailableCopy object becomes a
    BorrowedCopy object; Similar, after being returned, a BorrowedCopy object
    becomes available. Thus, a ReturnedCopy is a looped class of a AvailableCopy.
    It allows a trace of copy from the beginning of the loan to the end of the borrowing.

Figure 4  Example of loop situation in DS

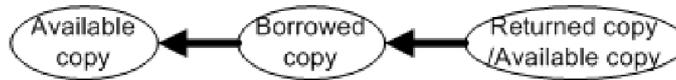

A DS may be imperative or non-imperative. In the special case that a sub-class has only one direct super-class (mono-specialisation), the relation must be imperative; in the case that a sub-class may have several direct super-classes (multi-specialisation) a sub-object must be related to at least one super-object.

### 4.2 Dynamic concepts

Dynamic concepts represent information processing. Two cases have to be distinguished: local view and global view. In the local view, the object life-cycle belongs to dynamic concepts. It describes changes of an object's state caused by process execution. Normally, a process corresponds to a business function or a decision. At a global view, dynamic concepts concern interaction between objects, business processes and information consumed or produced by processes that cause a change of the system's state. Our model takes into account both the local view and the global view.

In our model the object life-cycle is represented with the concepts of state and process. A process changes objects from input states into output states. The main differences between our model and others like UML state diagrams (Booch et al., 1998) and object life-cycles with Petri nets are as following:

- An object can remain in a state or leave it when it changes to another state.
    Thus, at any time an object may reside in several states. For example, when an
    AvailableCopy changes to a BorrowedCopy state, it abandons the AvailableCopy
    state; however, for instance, if a European country becomes member of the European
    Union, it would be necessary to keep its state as being a European country as well.

- A process is associated with execution rules that call pre- and post- conditions.
    A process can be executed if objects in its input states satisfy the pre-condition.
    After the process execution, its post-condition must be satisfied. These conditions
    are normally expressed as logical expressions using logical connectors (AND, OR,
    XOR) and combinations of them. The designer defines these conditions. There is no
    convention on logic connectors as in Petri nets.

### 4.3   Organisational concepts

The principal element in the organisational aspect is the concept of an organisational role. An organisational role corresponds to an organisational unit, which is assigned responsibilities and functions that should be carried out to achieve the objectives of the organisation. A process may have no assigned organisational role (responsibility of zero in case it is an automatic process) or one or several organisational roles. An organisational role may be responsible for zero, one or several processes. Organisational roles can have privileges on information accessibility such as creation, suppression, modification and query.

### 4.4   IASDO model definition

The formal IASDO model specification includes all concepts mentioned above, as well as the interrelation between these concepts. In the global view, we consider a state in the object's life-cycle of a class C as sub-class of class C. Thus, a state of C is viewed as a specific object of C that satisfies certain conditions. Furthermore, it is necessary that the order of state changes in the object's life-cycle must be consistent with the order of object migration specified in the DS. This ensures consistency between the two aspects. In this view, an object life-cycle of the dynamic aspects corresponds to a DS graph of the static aspects. Consequently, the input and output states of a process become information consumed and produced by the process with respect to execution rules.

Interrelations between the organisational aspect and the other two aspects are described by organisational roles that have responsibility for processes and privileges on information. This allows us to manage the process and the information in information manufacturing systems.

A complete specification of the IASDO model is defined as:

IASDO model = <CL, $f_{ed}$, $f_{ds}$, back-inactive, access-view, loop, P, $f_i$, $f_o$, R, Pr, $f_{rcl}$, $f_{rp}$>

- CL: a set of classes $\{cl_1, cl_2, \ldots, cl_n\}$.

    A class $cl_i$ = <Name, Att, Met>, $cl_i$.Att is the set of attributes of $cl_i$; $cl_i$.Met is the set of methods of $cl_i$.

- $f_{ed}$: an ED function, (CL $\times$ CL) $\rightarrow$ {0, 1, 2}.

    $f_{ed}$ ($cl_j$, $cl_i$) = 1 if there exists an imperative ED from the class $cl_j$ to the class $cl_i$; The value 2 indicates that the ED is non-imperative and the value 0 indicates the absence of ED from $cl_j$ to $cl_i$.

- $f_{ds}$: a DS function, (CL $\times$ CL) $\rightarrow$ {0, 1, 2}.

    $f_{ds}$ ($cl_j$, $cl_i$) = 1 if there exists an imperative DS link from the class $cl_j$ to the class $cl_i$, i.e., $cl_j$ is a direct sub-class of $cl_i$. The value 2 indicates that the DS link is non-imperative and the value 0 indicates the absence of DS from $cl_j$ to $cl_i$.

- back-inactive: (CL $\times$ CL) $\rightarrow$ {0, 1}, if back-inactive($cl_j$, $cl_i$) = 1 then when an object in class $cl_i$ becomes an instance of class $cl_j$ then the object becomes inactive in class $cl_i$ ($cl_i$ is an ancestor of $cl_j$).

- access-view: $CL \rightarrow \{CL.Att, CL.Met\}$,

  access-view$(cl_j) = \{\{cl_j.Att \cup cl_j.Accessible\_Att\}, \{cl_j.Met \cup cl_j.Accessible\_Met\}\}$

  where

  $cl_j.Accessible\_Att \subseteq \cup cl_i.Att)$ and $f_{ds}(cl_j, cl_i) = 1$, $i = 1..n$, $j = 1..n$, $i \neq j$

  $cl_j.Acessible\_Met \subseteq \cup cl_i.Met)$ and $f_{ds}(cl_j, cl_i) = 1$, $i = 1..n$, $j = 1..n$, $i \neq j$.

- loop: a function describes the end and start classes of a loop, loop : $CL \rightarrow CL$.

  if loop$(cl_j) = cl_i$, then $cl_j$ is the end of a loop which has its start at $cl_i$ ($cl_i$ is an ancestor of $cl_j$); $cl_j$ is called looped class of $cl_i$. This has the following consequence: when an object is created in $cl_j$, it is created again in $cl_i$.

- P: set of processes, a process has pre and post conditions.

- $f_i: (CL \times P) \rightarrow \{0, 1\}$, if $f_i(cl, p) = 1$ then cl is an input class of the process p.

- $f_o: (P \times CL) \rightarrow \{0, 1\}$, if $f_o(p, cl) = 1$ then cl is an output class of the process p.

- R: set of roles.

- Pr: set of privileges on information, Pr = {create, modify, delete, query}.

- $f_{rcl}$: assigns privileges on a class to a role, $(R \times CL \times Pr) \rightarrow \{0, 1\}$.

  if $f_{rcl}(r, cl, pr) = 1$ then the role r has the privilege pr on the class cl.

- $f_{rp}$: assigns responsibility for a process to a role, $(R \times P) \rightarrow \{0, 1\}$.

  if $f_{rp}(r, p) = 1$ then the role r is responsible for the process p, so a role r may be responsible for many processes and vice-versa.

A specification of the IASDO model also includes graphical specifications and a text description for detailed information about the attributes and methods of classes, active/inactive objects, access-view information and privileges information. In order to ensure a consistent specification between the three aspects, the model defines additional rules:

R1. An output class of a process has to link to at least one input class of this process by ED or DS link. This rule ensures the consistency between static and dynamic aspects: output information of a process cannot exist if there is no corresponding input information for this process.

R2. Let C1, C2 be two input classes of a process P; C3, C4 are output classes of the process P. If an organisational role R is responsible for P then R has at least the privilege of information creation on C3 and C4, the privilege of information consultation on C1 and C2.

## 5  Modelling information manufacturing systems with the IASDO model

In this section, we illustrate our model by modelling an Information Manufacturing System of a library management scenario (Pham Thi et al., 2006). Following we show how our model helps to increase the information quality by IMS modelling.

For our scenario we consider information related to Request, Loan, Reservation and state changes of a Copy. Figure 5 presents the information specification of the library management case. There is one DS graph whose root class is Copy and another DS graph whose root class is Request. Between Copy and Document is an ED relation. The first DS graph shows that an object in class Copy may become available copy and then this one may be blocked for a reservation or be borrowed, etc. When a copy becomes blocked, this copy is unavailable i.e., it is inactive in class AvailableCopy. Textual description of the specification is as follows.

- Back-inactive (BlockedCopy, AvailableCopy) = 1 means that an object that migrates from class AvailableCopy to BlockedCopy becomes inactive in AvailableCopy.

- Similarly, we have back-inactive (BorrowedCopy, AvailableCopy) = 1, back-inactive (NotifiedReservation, Reservation) = 1, etc.

- The access view of class AvailableCopy includes accessible attributes and methods such as access-view (AvailableCopy) = {available copy number, copy code, document number, available date, Borrow, Block}. The access view of class BorrowedCopy is access-view (BorrowedCopy) = {available copy number, copy code, document number, borrowing date, Return, Lose} etc.

The same applies for the DS graph (dotted classes) whose root class is Request. A request may become a reservation or a borrowing, before it is completed as finished borrowing. Textual description is Back-inactive (Borrowing, NotifiedReservation) = 1, Back-inactive (Borrowing, FinishedBorrowing) = 1, etc.

Figure 5   Information and relations between them in the library management case

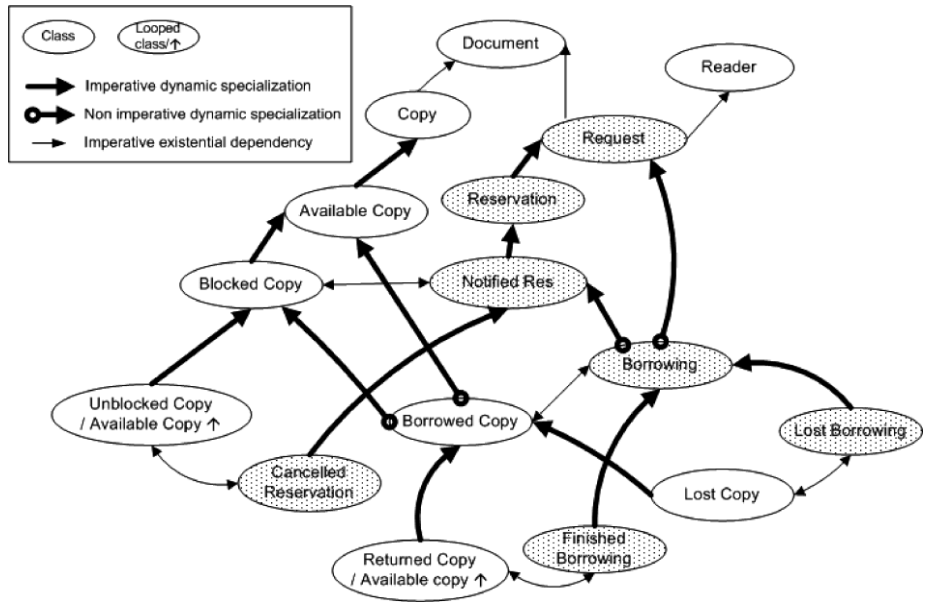

By using DS, the specification allows information about Copy and Document to be traced after each manufacturing step; relations between BorrowedCopy and Borrowing, BlockedCopy and NotifiedReservation, etc., allow a precise specification.

UnblockedCopy and ReturnedCopy are looped classes of the class AvailableCopy. For instance, an unblocked or returned copy becomes an available copy again.

The specification imposes rules on the order of state changes, e.g., it is impossible to change a borrowed copy to blocked copy because BorrowedCopy is a sub-class of BlockedCopy. Also, it imposes rules on the parentage of sub-objects, e.g., a borrowed copy of a loan must belong to the same document as the one requested by the corresponding loan. In contrast to the IASDO model, other modelling methods or models do not explicit include such rules.

Figure 6 illustrates information processing in the library management scenario. At a local view, the dotted part corresponds to the object life-cycle of Request. The remaining part (exclusion of class Document and class Reader) represents the object life-cycle of Copy. At a global view, this figure presents the information processing of IPs. A process consumes input information and produces output information, in accordance with its execution condition. The precondition of process Process request is (Request and Available Copy) or (Request); the post-condition is (Reservation) or (BorrowedCopy and Borrowing), etc.

Figure 6  Information processing in the library management case

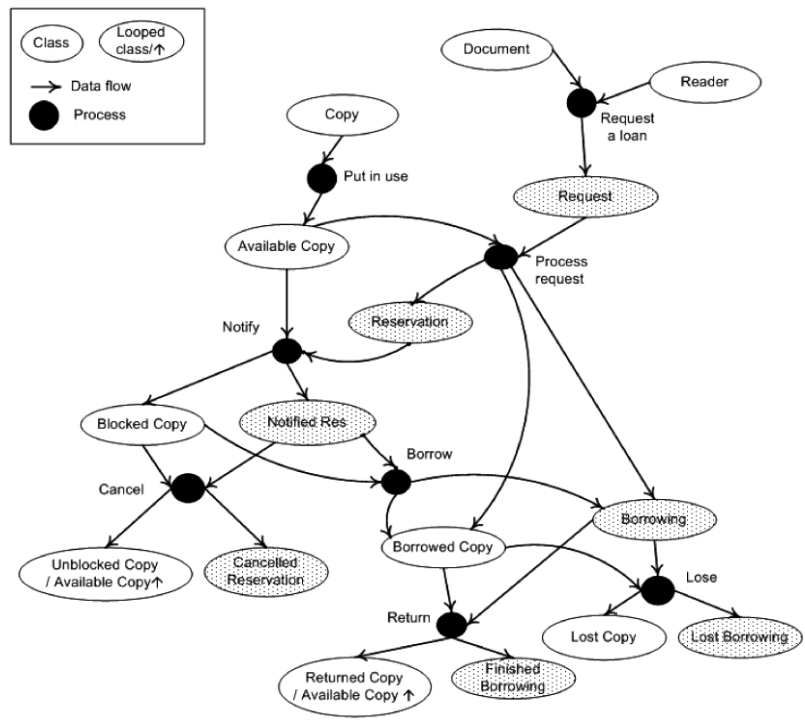

By integrating static and dynamic concepts, an object's life-cycle corresponds to a DS graph. Figure 7 illustrates the results. Organisational roles are described after each process' name. In this specification, Librarian, Logistic service and Control service are organisational units. For example, Librarian is responsible for Request a loan process, Borrow process, Notify process, etc., Librarian has creation privileges on Request, Loan, Reservation, BorrowedCopy, BlockedCopy, etc., and query or consultation privilege on AvailableCopy, etc.

Figure 7   Complete specification of information manufacturing of the library management case

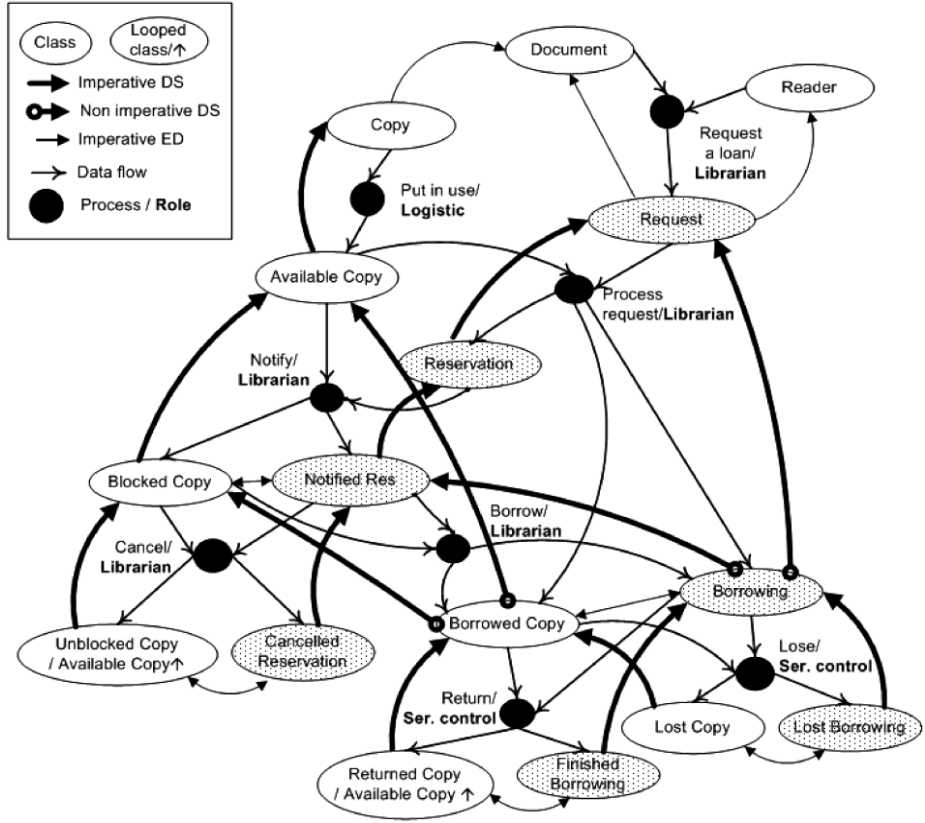

## 6   Quality evaluation

In the previous sections, we have presented the description of the IASDO model as well as indicated other modelling approaches. We have illustrated the application in a case scenario. Now we aim to compare the specification of the IASDO model with those of other methods and the models mentioned above. In summary, the IASDO model allows description of IPs, processes and organisational concepts. Furthermore, relations between IPs, as well as relations between IPs, processes and organisational roles are also explicitly described in this model.

DS allows us to describe each single information change while keeping track of information. The model supports designers to obtain a consistent specification. Execution rules are attached to processes; it provides a rigorous process description. Integration of organisational concepts with static and dynamic concepts allows us to describe organisational responsibilities and control of information access by describing explicit privileges on information. In addition, the model includes rules on cycles and validation rules. It assists designers to obtain a consistent and accurate specification.

Table 2 provides an overview of some selected modelling approaches, including our IASDO model. Our evaluation illustrates that the IASDO model might help to improve the information quality in information manufacturing systems.

Table 2  Quality evaluation of models for IMS modelling

| Criteria model | Completeness | Accuracy | Consistency |
| --- | --- | --- | --- |
| IASDO model | Modelling three aspects of IMS and interrelation between them | Automatically validate rules concerning cycles | DS supports the consistency between static and dynamic aspects |
|  | DS allows to keep track of information | Allows more precise semantic on data description | Additional proposed rules allow a consistent model regarding interrelation of three aspects |
| IP-Map | Modelling three aspects of IMS | Do not mention rules for guiding designer to obtain an accuracy model. It depends on experience of designers | Do not mention rules for guiding designer to obtain a consistent model. It depends on experience of designers |
| IP-UML | Modelling three aspects of IMS |  |  |
| EPC | Modelling three aspects of IMS |  |  |
| DFD | Modelling data and process aspect | Additional rules on the correctness including the accuracy of DFD | Additional rules on the correctness including the consistency of DFD |

7  Conclusions

Information quality in an Information Manufacturing System is among others affected by the quality of the conceptual model and the meta-model. The quality of the meta-model depends on how it supports the specification of the conceptual model, which one influences the quality of data. In this paper, we presented an overview of selected methods and models used for IMS modelling. In order to address some drawbacks of current approaches, we developed the IASDO model. The model includes static, dynamic and organisational aspects and helps to ensure consistency between the three aspects. The IASDO model improves the quality of the conceptual model due to its expressive power. It supports designers in obtaining an accurate and consistent model. For instance, it allows us not only to describe IPs, information manufacturing and organisational responsibility but also to describe interrelations between these aspects. The IASDO model includes rules, which allow us to validate the model. Furthermore, it allows trace-ability of information in the manufacturing process. Although our approach has advantages, it also increases the model's complexity, which might be seen as a limitation in adopting the approach in a practical setting. In our future research, we are aiming to address this by adopting our concept in a practical environment. We are implementing a prototypical tool based on the IASDO model approach. The tool can be used together with commercial available database systems and allows us to audit information processes in an information manufacturing system. Thus, the tool could support the cause analyses of inadequate information quality within an Information Quality Management Programme.

Acknowledgements

We would like to acknowledge Prof. Dr. Michel Léonard, Centre Universitaire d'Informatique, University of Geneva, Switzerland for his guidance and direction on the development of the IASDO model.

References

Al-Jadir, L. and Léonard, M. (1999) 'If we refuse the inheritance …', Proceeding DEXA, Florence, pp.560–572.

Ballou, D.P., Wang, R.Y., Pazer, H. and Tayi, G.K. (1998) 'Modelling information manufacturing systems to determine information product quality', Management Science, Vol. 44, No. 4, pp.462–484.

Bernus, P., Mertins, K. and Schmidt G. (1998) Handbook on Architectures of Information Systems, Springer-Verlag, ISBN: 3-540-64453-9.

Booch, G., Rumbaugh, J. and Jacobson, I. (1998) The Unified Modelling Language User Guide, Addison-Wesley object technologies series, Addison-Wesley.

Celko, J. (1987) 'I. data flow diagram', Computer Language, Vol. 4, January, pp.41–43. Demarco,

T. (1978) Structured Analysis and System Specification, Yourdon Press, New York. Duggan, E.W.

and Reichgelt, H. (2006) The Panorama of Information Systems Quality, Idea Group Publishing, ISBN: 1-59140-859-8.

Eppler, M.J., Helfert, M. and Gasser, U. (2004) 'Information quality: organisational, technological and legal perspectives', Studies in Communication Sciences, Vol. 4, No. 2, pp.1–16.

Keller, G., Nüttgens, M. and Scheer, A-W. (1992) Semantische Prozeßmodellierung auf der Grundlage Ereignisgesteuerter Prozeßketten (EPK), Technical Report 89. Institut für Wirtschaftsinformatik Saarbrücken. Saarbrücken, Germeny.

Krogstie, J., Lindland, O.I. and Sindre, G. (1995) 'Defining quality aspects for conceptual models', Proceedings of the International Conference on Information System Concepts (ISCO3), Towards a Consolidation of Views, Marburg, pp.216–231.

Léonard, M. (1999) 'M7: une approche évolutive des systèmes d'information', Proceeding INFORSID, La Garde, France (in French), pp.9–28.

Lindland, O.I., Sindre, G. and Solvberg, A. (1994) 'Understanding quality in conceptual modelling', IEEE SOFTWARE, Vol. 2, pp.42–49.

Moody, D.L. (2005) 'Theoretical and practical issues in evaluating the quality of conceptual models: current state and future directions', Data and Knowledge Engineering Journal, Vol. 55, pp.243–276.

Moody, D.L. and Shanks, G. (1994) 'What makes a good data model? Evaluating the quality of entity relationship models', The 13rd International Conference on entity Relationship Approach, LNCS, Vol. 881, pp.94–111.

Pham Thi, T.T. (2005) Intégration des Aspects Statique, Dynamique et Organisationnel dans la Modélisation des Systèmes D'information, PhD Dissertation, University of Geneva (in French).

Pham Thi, T.T. and Leonard, M. (2006) 'An advanced data model for information systems modelling', The 2nd International Advanced Databases Conference, June, San Diego, USA, ISBN: 0-9742448-6-4.

Pham Thi, T.T., Helfert, M. and Duncan, H. (2006) 'The IASDO model for information manufacturing systems modelling', The 11th International Conference on Information Quality (ICIQ'06), Massachusetts Institute of Technology, USA.

Pipino, L., Lee, Y.W. and Wang, R.Y. (2002) 'Data quality assessment', Communication of ACM, Vol. 45, No. 4, pp.211–218.

Scannapieco, M., Pernici, B. and Pierce, E. (2005) 'IP-UML a methodology for quality improvement based on information product maps and unified modelling language', in Wang, R.Y., Madnick, S.E. and Pierce, E.M. (Eds.): Information Quality. Advances in Management Information Systems, Vol. 1, M.E Sharpe Publisher, ISBN 0-7656-1133-3, pp.115–131.

Schuette, R. and Rotthowe, T. (1998) 'The guidelines of modelling – an approach to enhance the quality in information models', International Conference on Conceptual Modelling, ER'1998, LNCS, Vol. 1507, pp.240–254.

Shankaranarayanan, G., Wang, R.Y. and Ziad, M. (2000) 'IP-MAP: representing the manufacture of an information product', Proceeding of the 6th International Conference on Information Quality, Massachusetts Institute of Technology, USA.

Snoeck, M. and Dedene, G. (1998) 'Existence dependency: the key to semantic integrity between structural and behavioural aspects of object types', IEEE Transaction on Software Engineering, Vol. 24, No. 4, April, pp.233–251.